\documentclass[12pt,preprint]{aastex}

\usepackage{natbib}
\usepackage{graphicx}
\usepackage{amssymb}

\shortauthors{Yelles Chaouche et al.}
\shorttitle{Mesogranulation and the solar surface magnetic field distribution}

\def\rhocork{{\rho_{c}^{}}}
\def\rhocorkzero{{\rhocork_0^{}}}
\def\Bzthr{{|B_z|}_{th}^{}}

\begin{document}

\title{Mesogranulation and the solar surface magnetic field distribution }

   \author{L. Yelles Chaouche\altaffilmark{1,2}, F. Moreno-Insertis\altaffilmark{1,2}, V.
   Mart\'inez Pillet\altaffilmark{1}, T. Wiegelmann\altaffilmark{3}, J. A. Bonet \altaffilmark{1,2}, 
   M. Kn\"olker\altaffilmark{4,5}, 
   L. R. Bellot Rubio\altaffilmark{6}, J.C. del Toro Iniesta\altaffilmark{6}, 
   P. Barthol\altaffilmark{3}, A. Gandorfer\altaffilmark{3}, W. Schmidt\altaffilmark{5}, 
   S.K. Solanki\altaffilmark{3,7}}

\altaffiltext{1}{Instituto de Astrofisica de Canarias, Via Lactea, s/n,
  38205  La Laguna (Tenerife), Spain}
\altaffiltext{2}{ Dept.~of Astrophysics, Universidad de La
  Laguna, 38200 La Laguna (Tenerife), Spain}

\altaffiltext{3}{Max-Planck-Institut f\"ur Sonnensystemforschung,
Max-Planck-Strasse 2, 37191 Katlenburg-Lindau, Germany}

\altaffiltext{4}{High Altitude Observatory (NCAR), Boulder, CO 80307, USA}

\altaffiltext{5}{Kiepenheuer-Institut f\"ur Sonnenphysik, 79104, 
Freiburg, Germany}

\altaffiltext{6}{Instituto de Astrofisica de Andalucia (CSIC),
Glorieta de la Astronomia, s/n
18008 Granada, Spain}

\altaffiltext{7}{School of Space Research, Kyung Hee University, Yongin,
Gyeonggi, 446-701, Korea}

\date{}

\begin{abstract}

  The relation of the solar surface magnetic field with mesogranular cells is
  studied using high spatial ($\approx 100$ km) and
  temporal ($\approx 30$ sec) resolution data obtained with the IMaX instrument
  aboard SUNRISE. First, mesogranular cells are identified
  using Lagrange tracers ({\it corks}) based on horizontal velocity fields
  obtained through Local Correlation Tracking.  After $\approx 20$ min of
  integration, the tracers delineate a sharp mesogranular network with
  lanes of width below about $280$ km. The preferential location of
  magnetic elements in mesogranular cells is tested
  quantitatively. Roughly $85\%$ of pixels with magnetic field higher than
  $100$ G are located in the near neighborhood of mesogranular lanes.
  Magnetic flux is therefore concentrated in mesogranular lanes rather than
  intergranular ones.

  Secondly, magnetic field extrapolations are performed to obtain
  field lines anchored in the observed flux
  elements. This analysis, therefore, is independent of the
  horizontal flows determined in the first part. A
  probability density function (PDF) is calculated for the distribution of
  distances between the footpoints of individual magnetic field lines.
  The PDF has an exponential shape
  at scales between $1$ and $10$ Mm,  with a constant
 characteristic decay distance, indicating the absence of preferred
 convection scales in the mesogranular range.
  Our results support the view that
  mesogranulation is not an intrinsic convective scale (in the sense that
  it is not a primary energy-injection scale of solar
  convection),  but also give quantitative confirmation that,
    nevertheless, the magnetic elements are preferentially found 
    along mesogranular lanes.
  \end{abstract}


\section{Introduction}\label{sec:introduction}
    
Mesogranulation was historically introduced as a prominent scale imprinted on
the horizontal photospheric flows calculated through local correlation tracking
(LCT) of intensity images \citep{november1981, simon1988,
  brandt1988, brandt1991, muller1992, roudier1998, roudier1999, shine2000,
  leitzinger2005}: both the pattern of positive and negative divergence of
that flow as well as the time evolution of Lagrange tracers moving in it
revealed cells with sizes between, say, $5$ and $10$ arcsec. Much debate
ensued concerning whether the mesogranular flow patterns correspond to actual
convection cells in that range of sizes rather than, e.g., to simple granule
associations which persist in time \citep[see, e.g.][]{cattaneo2001,
  roudier2003, roudier2004, nordlund2009, matloch2009, matloch2010}.
Independent hints for the existence of a convective flow operating on those
scales are therefore of importance. The study of the surface distribution of
magnetic elements can provide such hints; as a minimum, it can constitute an
alternative avenue, independent of the inaccuracies of LCT methods, to
determine the properties of the mesogranular patterns. Such an approach
has been employed by \citet{dominguezetal2003, dominguez2003,
  sanchez_almeida2003} using ground-based data and by \citet{roudier2009} and
\citet{ishikawa2010} using satellite observations. In those
papers, visual evidence was obtained that there is an association between
magnetic flux structures (flux elements, transient horizontal fields) and the
mesogranular pattern obtained through the study of horizontal flows. Yet,
detailed quantitative studies and statistics that could put such an
association on a firmer basis are still missing.

The aim of this letter is to obtain quantitative information of the relation
between photospheric magnetic flux distributions and mesogranular scales by
using the observations of unprecedented quality provided by the Imaging
Magnetograph eXperiment (IMaX; \citealt{martinezpillet_etal_2010}) aboard the
SUNRISE balloon-borne observatory \citep{barthol_etal_2010, solanki_etal_2010}. IMaX provides time series of
virtually seeing-free high spatial resolution ($0.15$ arcsec) images and
magnetograms that constitute an ideal data set for the study of
photospheric magnetism.  
Using the Sunrise/IMaX data, we combine
information from (a) the velocity field gained through Local
Correlation Tracking of intensity images; (b) the spatial patterns
provided by the magnetograms; and (c) the field line structure obtained through extrapolation
of the magnetogram data, to gain quantitative information concerning
patterns at intermediate scales between granulation and supergranulation.

\section {Data}\label{sec:data}

\begin{figure} 
\includegraphics[width=0.5\textwidth]{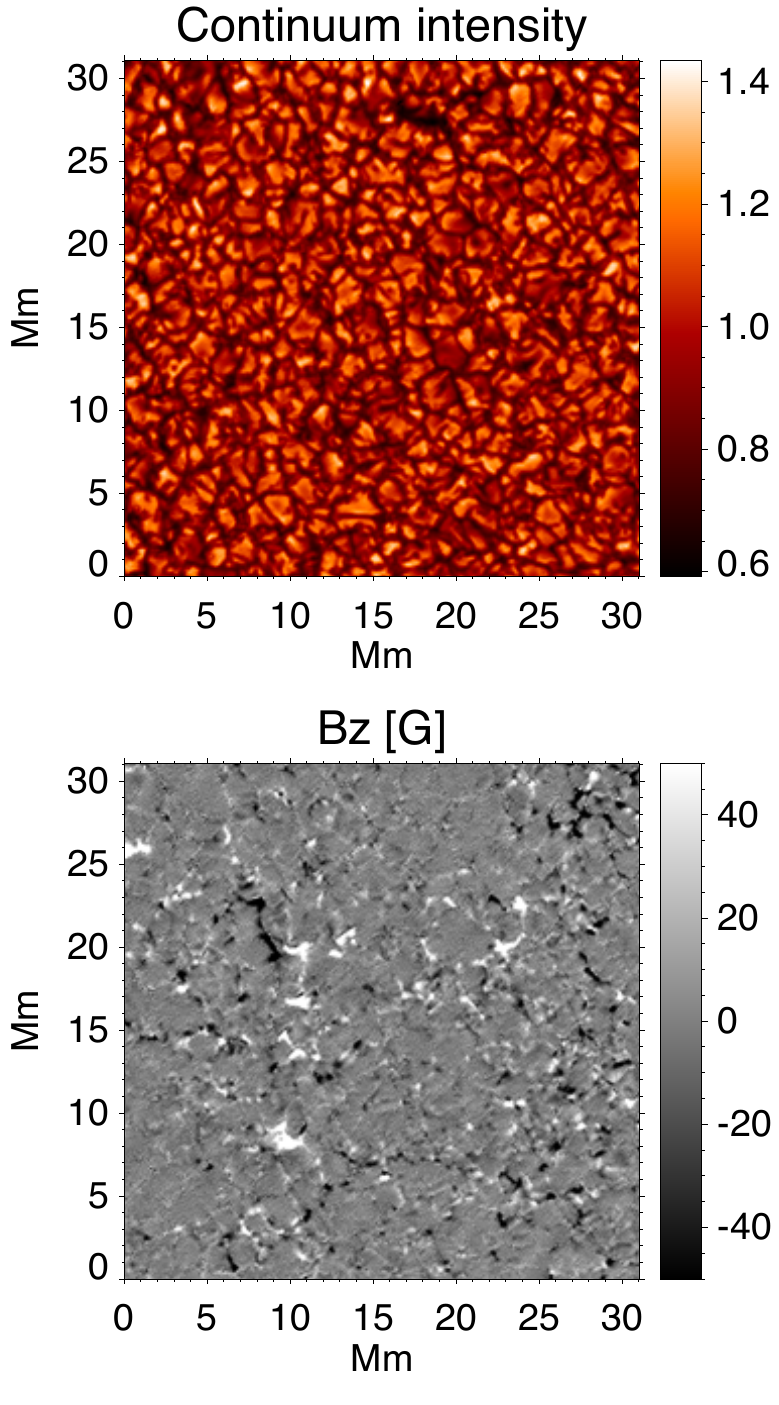}

\caption{Top: normalized continuum intensity near the spectral line Fe I
  5250.2 \AA. Bottom: Vertical component of the magnetic field vector ($B_z$)
  retrieved from inversions and clipped at $\pm 50$ G. The maps are taken from the first observed time
  series.} \label{fig1}
\end{figure}

For this study we use sequences of images recorded with IMaX near the solar disk center on 2009 
June 9. Images were taken at five wavelengths along the profile of the
magnetic-sensitive FeI $5250.2$ \AA\ line located at $\pm 80$ m\AA, $\pm 40$ m\AA\ from line
center, and continuum at $+227$ m\AA. The estimated circular polarization noise is $5\times 10^{-4}$ in 
units of the continuum wavelength for non-reconstructed data and $3$ times larger for the reconstructed one. 
IMaX has a  spectral resolution of $85$ m\AA\ and a spatial
resolution of $\sim 100$ km \citep{martinezpillet_etal_2010}. The reduction procedure produces time series of 
images with a cadence of $33.25$ s, 
spatial sampling of $39.9$ km 
and a field-of-view (FOV) of $32 \times 32$ Mm$^2$. 
We use two time series, the first one comprising $42$ 
snapshots ($23$ minutes) 
and the second one $58$ snapshots ($32$ minutes). 
Magnetograms are derived from inversions
of the observed Stokes parameters using the SIR code  
\citep{ruiz_cobo1992}. We call these magnetograms the {\it
  reconstructed data}. In this letter we mainly use the reconstructed data
except in the last section (Sec.~\ref{sec:extrapolation}) where, in addition,
we use so-called {\it calibrated data}, i.e., magnetograms obtained using a
proportionality law between Stokes-V (from non-reconstructed data) and $B_z$
\citep[for details of this method, see][]{martinezpillet_etal_2010}. The intensity 
maps are taken from the reconstructed data.

Figure~\ref{fig1}
displays a map of normalized continuum 
intensity (top) and the corresponding longitudinal magnetogram (bottom), the latter showing 
many internetwork flux concentrations alongside stronger flux elements probably belonging 
to the network.

\section{Mesogranulation and horizontal flows}\label{sec:mesogranulation}

Following the traditional method, mesogranular patterns are outlined
by using Lagrange tracers (popularly known as {\it corks}) arranged at time
$t=0$ uniformly in a 2D grid that coincides with the observational pixel
matrix.  Then, the corks are advected following the instantaneous horizontal
velocity field along the whole duration of the time series (here we use the 
second time series: $32$ minutes). The horizontal
velocity field is determined using the local correlation tracking (LCT)
algorithm by \citet{welsh2004} applied to consecutive continuum intensity images. 
The correlation is performed in local windows weighted by
a Gaussian function with FWHM=320 km. Often, in the literature, the 
velocity field determined through LCT and used to advect corks is a time average of the noisy determinations of the instantaneous velocity 
fields for consecutive snapshots \citep[for a discussion about the effect of time averaging see 
e.g.][]{rieutord2000, rieutord2001}.
The high signal-to-noise ratio, the high spatial and 
temporal resolution and the absence of atmospheric
distortion in the IMaX data make it possible to evaluate clean velocity
fields and consequently to compute cork advection without time averaging.

In Fig.~\ref{fig5} we show (white dots) the distribution of corks at $t=28$ min, i.e.,
toward the end of the series, on the background of the continuum intensity
image at that time. The corks clearly delineate cells of mesogranular size
containing several granules. In fact, the vast majority of corks are
located in the lanes between those cells (which we will call here
mesogranular lanes or {\it mesolanes} for short).  A clear mesogranular
pattern with the majority of the corks concentrated in mesolanes is already
visible after only some 20 minutes from $t=0$. This is less than the time
(2-4 hours) reported by \cite{roudier2009} using the Hinode/SOT NarrowBand
Filter Imager.

In order to obtain quantitative measures for the mesogranular network we
define a {\it cork density function}, $\rho_{cork}$, by counting the number
of corks in each pixel in the image. 
To provide an upper bound for the width of the mesolanes, we scan the
image vertically using whole horizontal lines and calculate the width 
($1/e$ of the maximum) of the peaks of the $\rho_{cork}$ function in each line. We then derive an
average peak width for the whole scan. If we use all peaks above $\rho_{cork}
= 10$, then the resulting upper bound for the lane width is $285$ km. The largest concentrations of
corks are located in narrow mesolanes: using all peaks above
$\rho_{cork}=30$, for instance, yields a width of $180$ km ($4.5$
pixels). Given the random orientation of the mesolanes, their actual width is
certainly below those numbers.  Using a horizontal scan instead of a vertical
one yields basically the same numbers (the maximum deviation is $7\%$).

\section{Correlation between mesogranular lanes 
and field concentrations} 

A first test of the relation between mesogranules and the surface magnetic
field can be obtained by studying the spatial association between
mesogranular cells and magnetic flux concentrations. Through visual
inspection, various researchers have obtained
indications that the magnetic field concentrations with higher flux density
are distributed at the boundaries of cells of mesogranular size ($5$ to $10$
arc sec; see \citealt{dominguezetal2003} and \citealt{sanchez_almeida2003});
in fact, \citet{dominguez2003} shows the preference of elements with flux
density above $60$ G to be located in regions of high negative divergence
of the horizontal vlocity. Further visual evidence is provided by
\citet{roudier2009} who plot Stokes-V images obtained with Hinode/NFI on top
of cork distributions showing a relationship between magnetic field
concentrations and cork lanes \citep[see also][]{de_wijn_etal_2008, solanki_etal_2010}.

\begin{figure} 
\begin{center}

 \includegraphics[width=0.50\textwidth]{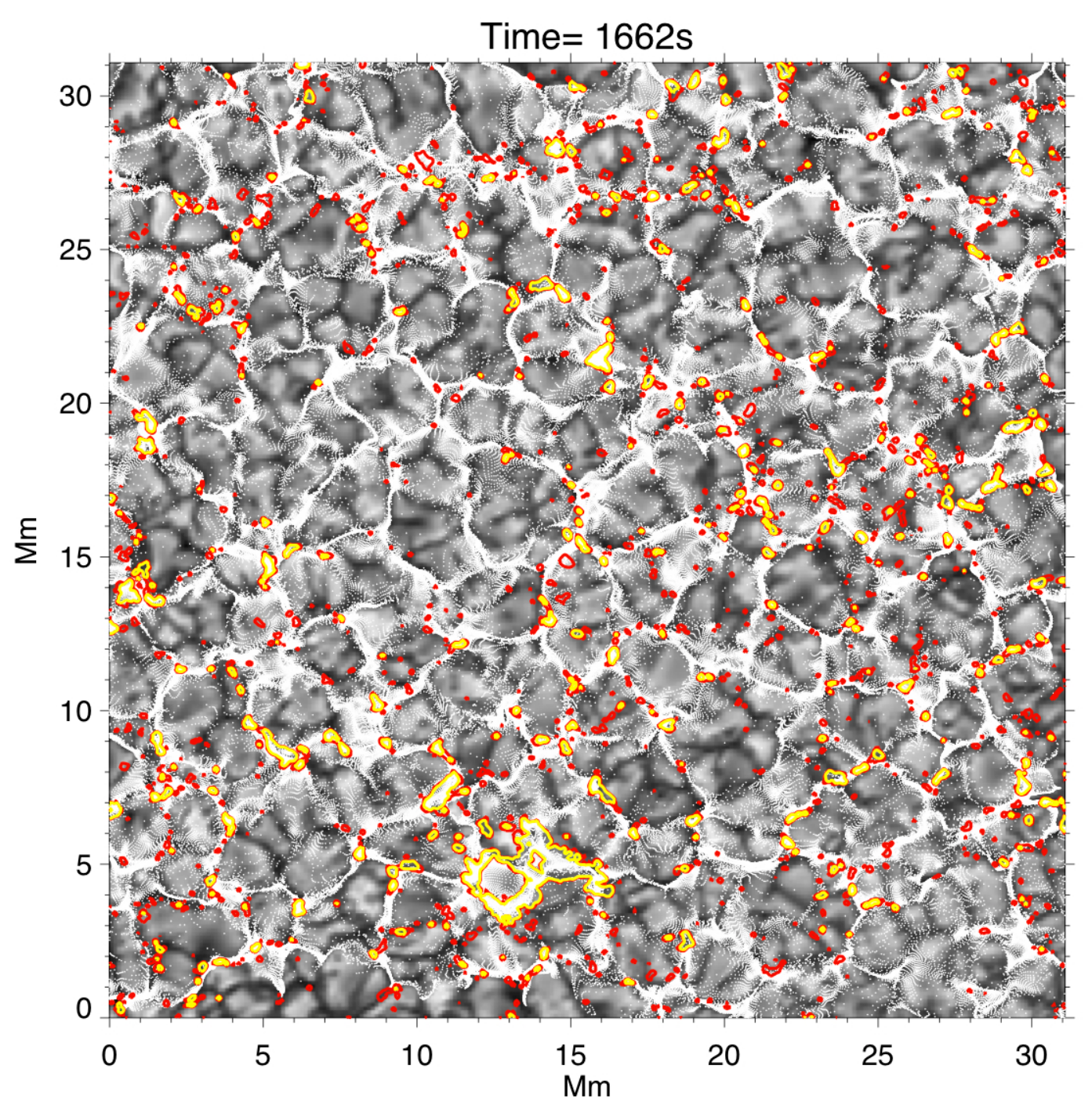}
\end{center}
\caption{ Cork distribution (white dots) outlining mesogranular cells at t=28
  min. The background image is continuum intensity. Color lines: $|B_z|$
  contours for $30$ G (red) and $50$ G (yellow).} \label{fig5}
\end{figure}

The conclusions at that level can be reinforced through our
high-resolution maps by drawing contours of $|B_z|$ down to small values,
like $30$ G. The red and yellow contours in
Fig.~\ref{fig5} correspond to flux density values of $30$
and $50$ Gauss, respectively. The figure exhibits flux
concentrations mostly located at mesolanes.
One can notice that flux concentrations with
higher flux density (yellow contours: $50$ G) correlate better with
mesolanes. 

\begin{figure} 
\begin{center}

\includegraphics[width=0.5\textwidth]{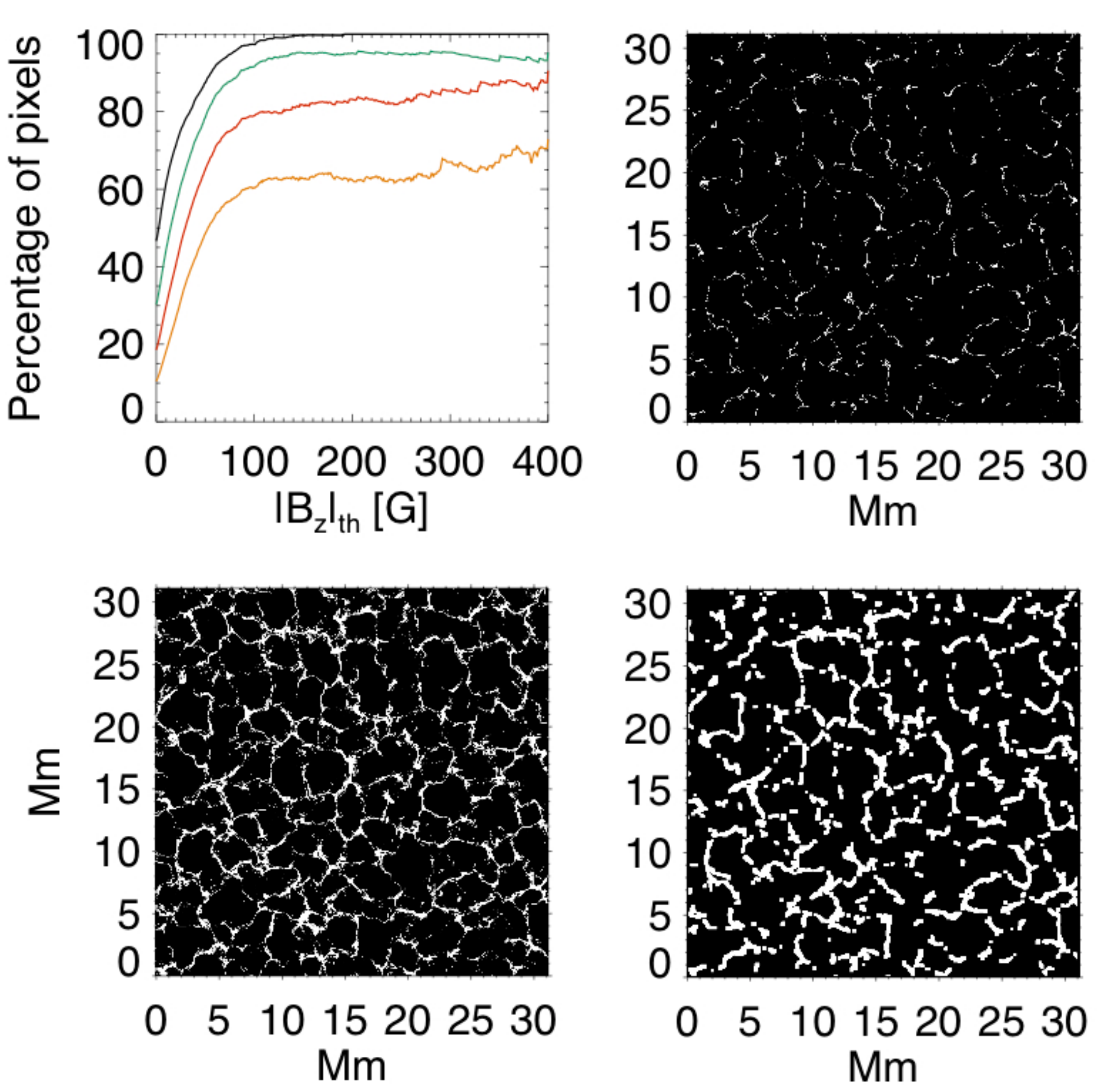}
\end{center}
\caption{Top-left: percentage of pixels with $|B_z|_{th}$ above the value
  given in abscissas located within 3 pixels ($\sim$120 Km) 
  from mesolanes. Mesolanes are defined through $\rhocork > 2, 4, 8,
  16$ corks/pixel for the black, green, red and orange curves, respectively. 
  Top-right and bottom-left panels: distribution of pixels with $\rhocork >
  8$ (top right) and $\rhocork > 2$ (bottom left) corks/pixel, respectively. Bottom-right panel:
  3-pixel neighborhood of the points shown in the top-right panel. The calculations
  are done for similar time as Fig.~\ref{fig5} (t=28 min).
} \label{fig6} 
\end{figure}

Going one step further, we can obtain quantitative estimates to test the
visual impression gained by combining cork images and magnetograms.  We use
as a proxy for the mesolanes the locations with cork density above a
threshold, $\rhocork > \rhocorkzero$, and use the values $\rhocorkzero = 2,
4, 8, 16$ corks / pixel. All of those values yield clear mesogranular lanes,
as apparent, e.g., in the top-right and bottom-left panel of Fig.~\ref{fig6},
drawn for $\rhocorkzero =8$ and $2$, respectively. Increasing $\rhocorkzero$
decreases the level of connectivity of the lanes. We then focus on the
mesolanes defined by $\rhocork > \rhocorkzero=8$ and study which fraction of
the magnetic elements are really in (or near) them. To obtain a quantitative
estimate, for each fixed threshold intensity $\Bzthr$ in the range $(0,\,400)$
G, we consider the set of pixels with $B>\Bzthr$ and calculate which fraction
of the set is located in 3-pixel neighborhoods of the mesolanes. The result
is shown in the top left panel of Fig.~\ref{fig6}, red curve. The curve
reaches an approximate horizontal asymptote at the $85\%$ level for $\Bzthr$
above $100$ G.  The other curves in the figure contain the
same results but using $\rhocorkzero = 2$ (black curve), $4$ (green) and $16$
(orange). We see approximate horizontal asymptotes in all cases, and all are
reached for $\Bzthr$ between $80$ to $100$ G. We conclude that the vast
majority of pixel elements with $B \gtrsim 100$ G is located in the near
neighborhood of locations with high cork density. For the least restrictive
case ($\rhocorkzero=2$), virtually all magnetic elements with $B \gtrsim
100$G are located in the neighborhood of the mesolanes.  For large
$\rhocorkzero$ (e.g., $\rhocorkzero=16$ c/p, orange curve), the resulting
areas have less connectivity and do not delineate so clearly the mesogranular
network. Correspondingly, the asymptotic percentage values for $B\gtrsim 100$
G become smaller.  The width of the 3-pixel neighborhoods of the mesolanes is
shown in the bottom-right panel for the $\rhocorkzero = 8$ corks/pixel
case. The panel shows a fully developed network of mesolanes covering only
$17$\% of the whole surface (which, of course, coincides with the $\Bzthr \to
0$ limit of the red curve).  As an aside note, all curves remain basically
the same if we remove the prominent network element located in the lower part of
Fig.~\ref{fig5}.

\section{The separation between the footpoints of extrapolated field lines}
\label{sec:extrapolation}

\subsection{Method}
As a further test of the relation between surface fields and convection at
different scales, in this section we study the statistics of footpoint
separation between field lines linking the magnetic elements observed with
IMaX.  To that end, we calculate the magnetic field vector in
the atmosphere using force-free field extrapolations (FFF) from the IMaX data. 
We use a code with weighted
optimization method (\citealt{ wiegelmann2004}; see also
\citealt{seehafer1978}). The force-free assumption is equivalent to assuming
that the electrical current $\mathbf{j}$ and the magnetic field $\mathbf{B}$
are parallel, or, using Amp\`ere's law:
\begin{eqnarray}
  \nabla\times \mathbf{B} = \frac{4\,\pi}{c}\mathbf{j} = \alpha\, \mathbf{B} \label{eqn:exp1}\;;
\end{eqnarray} 
$\alpha$ in Eq.~\ref{eqn:exp1} measures the level of field line twist. 
The field values obtained with the IMaX data correspond to
the height where the Fe I 5250.2 \AA\ line is formed. 
Although the
photosphere is not a small-$\beta$ plasma (with $\beta$ the ratio of gas
to magnetic pressures), there are theoretical and observational
indications \citep{wiegelmann2010a, wiegelmann2010b, martinez_gonzalez2010}
that the force-free assumption is acceptable when calculating
extrapolations starting from those heights. We will be using $\alpha = 0.1 /
L$, with $L \times L$ being the IMaX field of view.  In any case, \citet{wiegelmann2010b},
using the same IMaX data, show that values of $\alpha$ in the range $(-4/L, 4/L)$
lead to similar values in the statistical properties they analyzed for the
extrapolated field lines. 

For computational reasons, we have calculated the extrapolations in a box of
$389$ $\times$ $389$ $\times$ $389$ equally spaced grid points, keeping the original
horizontal dimensions, leading to a cell size of $\approx 80$ km. Yet,
test calculations with the original number of grid points ($778$ in each
direction) yield essentially the same results.

\begin{figure} 
\begin{center}
\includegraphics[width=0.35\textwidth]{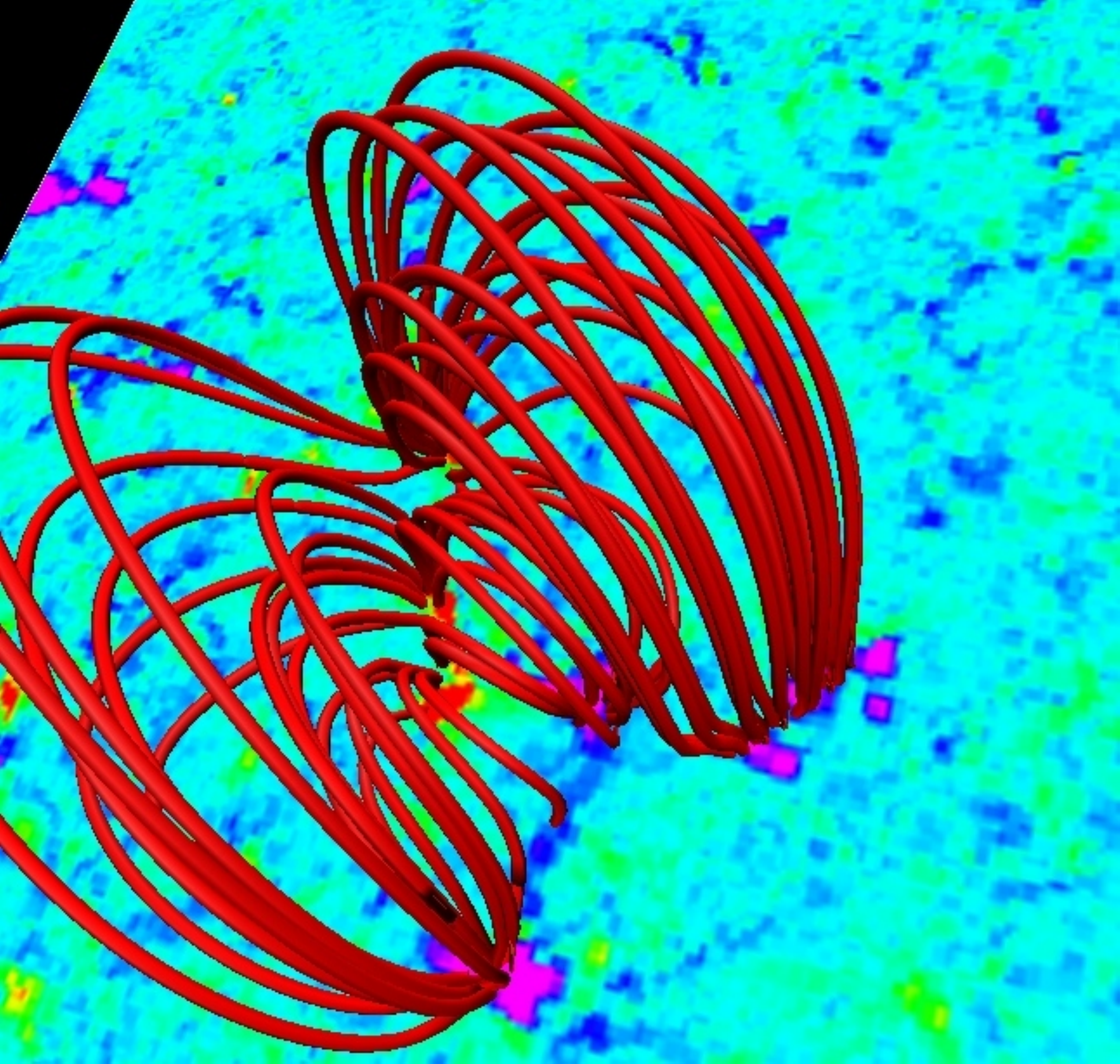}
\end{center}
\caption{ Magnetic field lines showing the connectivity between flux
  concentrations. Background image: map of $B_z$ (purple and blue: $B_z>0$;
  green and red:  $B_z<0$).  
  The background image is a $\approx$ $10 \times 10 $ Mm$^{2}$ patch around
  the point $(10, 20)$ Mm in Fig.~\ref{fig1}.} \label{fig3}
\end{figure}

An example of magnetic field lines is shown in Fig.~\ref{fig3}. A map of the
vertical magnetic field component is shown at the bottom of the box
corresponding to a region of about $10\times 10$ Mm$^2$ centered on the point
$(x,y) = (10, 20)$ Mm in Fig.~\ref{fig1}. The field line footpoints are seen
to cluster on field concentrations in the domain. In
agreement with the conclusions of \cite{wiegelmann2010b}, this
figure shows a rough proportionality between the footpoint distance
and the height reached by the field lines.

\subsection{Statistics of footpoint separation}

\begin{figure} 
\begin{center}

\includegraphics[width=0.5\textwidth]{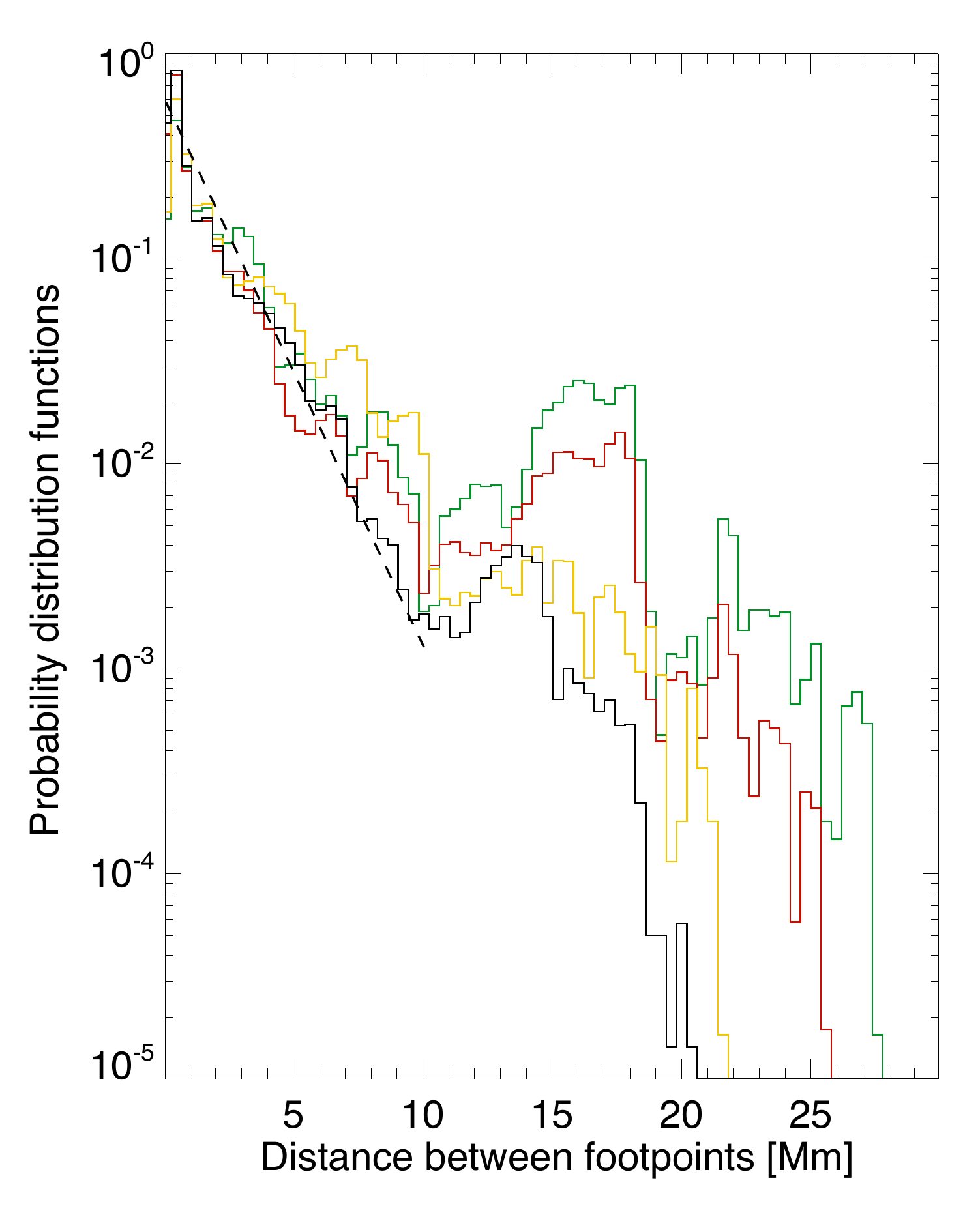}

\end{center}
\caption{Probability density functions of the distance between footpoints of
  field lines. Black and yellow: data from the first time series (inverted
  data: black; data calibrated following
\citet{martinezpillet_etal_2010}: yellow). Red and green: same as black and
  yellow, but for series 2.  
  The straight dashed line represents an exponential fit of the black curve for the first $\approx$10
  Mm.} \label{fig4} 
\end{figure}

Calling $x$ the distance between footpoints, we calculate the probability
density function (PDF) for $x$ using as statistical ensemble for each time
series the field lines in each snapshot and the whole collection of snapshots
in the series.  To include a field line in the study, we request a flux
density above $15$ G on both footpoints. The resulting PDFs are shown in
Fig.~\ref{fig4}.  The black and red lines correspond to the first and second
time series, respectively. Additionally, we plot PDFs for distances between
field line footpoints calculated by extrapolating calibrated data following
\citet{martinezpillet_etal_2010}, as explained in Sec.~\ref{sec:data}, which
involves no inversion procedure (yellow and green curves, for the first and
second time series, respectively).  All the curves approximately fit an
exponential distribution at scales between $1$ and $10$ Mm, i.e., at
granular and mesogranular scales; the slope is $d \log \hbox{(PDF)} /dx$ $\approx$
$-0.6$ Mm$^{-1}$.  Beyond $10$ Mm, at supergranular scales, the curves show a
deviation from the exponential.  This {\it bump} most probably
corresponds to the presence in that time series of strong network elements. 
Anyway, given
the size of the IMaX field of view ($\approx 30$ Mm) and the long duration of
the largest convection cells, we cannot draw statistical inferences from the
PDF for scales above, say, $10$ Mm. On the other hand, we have carried out a Kolmogorov-Smirnov
test of the goodness of fit of the footpoint distance data in the $1$-$10$ Mm
range to a lognormal distribution, additionally to the exponential
distribution; a lognormal distribution is found to fit well magnetogram data
or, more generally, data resulting from the fragmentation of magnetic
elements (\citealt{Abramenko2005}; see also \citealt{Bogdan1988}). In our
case, the test, carried out for individual snapshots in the given distance
range, favors the exponential distribution. Details of this analysis will be
given in a publication in preparation.

We note the constancy of the
slope of the PDF at scales between $1$ and $10$ Mm
and that the characteristic decay
distance is approximately $1.7$ Mm. This is
probably a consequence of the fact that there are no intrinsic horizontal
scales in that range other than granulation, e.g., because mesogranulation
is the direct result of other convection scales,  rather than representing a
primary scale in which energy is being injected into the convective flows.

\section{Discussion and Conclusions}

An important open question in solar physics is the precise nature of the
convection scales with size and duration above the granular values. While
both granulation and supergranulation yield velocity field patterns that have
been observed at the surface, mesogranules have been detected only through
indirect proxies, like tracking of intensity patterns, which do not provide
reliable evidence of underlying convection cells in that range of sizes and
durations. There is an ongoing debate on whether there is a continuum of
sizes for the convection cells on scales above granular, possibly with
self-similar properties and with no particular scale being singled out within
that range \citep{nordlund2009}, or whether the mesogranular scales are just
the result of a collective interaction between families of granules. The
magnetic field can provide an alternative and more direct avenue to explore
convective patterns since the magnetic flux can be measured directly using
Stokes polarimetry techniques. The magnetic elements appear with a broad
spectrum of flux densities
\citep{orozco_suarez_etal_hinode_magnetic_elements_2007, khomenko_etal_2003}
and can also be detected through G-Band or Ca-II bright points
\citep[see][]{de_wijn_etal_2008, sanchezalmeida_etal_2010} and
  therefore can provide important clues concerning the nature of the flows
  underlying the mesogranular scales as well as about the magnetic elements themselves. 

The high spatial and temporal cadences of the IMaX data allow us
to try a few different, complementary studies of the relation between
magnetic field and mesogranular flows. First, we have carried out Lagrange tracing
of mass elements following horizontal flow fields obtained through LCT of
intensity maps; we have obtained a number of improvements
compared with traditional cork maps (faster development of the mesogranular
lanes, no need for time averages). We have also obtained an upper bound for
the width of the mesogranular lanes of some $280$ km.  Second, we have
provided quantitative measures for the association between magnetic elements
and mesogranular lanes. The large majority ($85$ \%) of the magnetic
elements with flux density above $\sim 100$ G are found in $120$-km
neighborhoods  of mesogranular lanes with $\rhocork > 8$ and about $80$ \% 
of flux elements above $30$ G are located near mesogranular lanes with $\rhocork > 2$.  
 Our results indicate
  a good coupling between the flow field and the magnetic elements,
  suggesting that the evolution of the latter is mostly kinematic. 
Third, we have considered the
connectivity between magnetic elements and, in particular, the distance
between footpoints of field lines anchored in the photosphere.  The
probability density function of such distances shows that there is abundant
connectivity on mesogranular scales; it also shows that the distribution is
basically featureless in those scales, with only one characteristic value,
the slope of the distribution, equal to $(1.7\, \hbox{Mm})^{-1}$.

Our results concerning statistics of separations of field line footpoints
suggest that there is no intrinsic scale of convection in the mesogranular
range. This may only mean that there is no mechanism for direct injection of
energy into convection on those scales, but does not rule out the existence
of convection cells with those sizes, e.g., through nonlinear interactions of
cells at other scales (as in a turbulent cascade) or through the interaction
of thermal downflows \citep{rast2003}. 


\vskip 5mm

\acknowledgements{
  Financial support by the European Commission through the SOLAIRE Network
  (MTRN-CT-2006-035484) and by the Spanish Ministry of Research and
  Innovation through projects AYA2007-66502, CSD2007-00050 and AYA2007-63881 is gratefully
  acknowledged, as are the computer resources, technical expertise and
  assistance provided by the MareNostrum (BSC/CNS, Spain) supercomputer
  installation. L.Y.C. and F.M-I. are grateful to V. Abramenko for advice
  concerning the Kolmogorov-Smirnov test of statistical distributions. 
The German contribution to Sunrise is funded by the Bundesministerium f\"ur
Wirtschaft und Technologie through DLR Grant 50 OU 0401, and by the
Innovationsfond of the President of the Max Planck Society (MPG). The Spanish
contribution has been funded by the Spanish 
MICINN under projects ESP2006-13030-C06 and AYA2009-14105-C06. The HAO
contribution was partly funded through NASA grant number NNX08AH38G.
This work has been partially supported by WCU grant No. R31-10016 funded by the
Korean Ministry of Education, Science and Technology.   
}


\begin{thebibliography}{39}
\expandafter\ifx\csname natexlab\endcsname\relax\def\natexlab#1{#1}\fi

\bibitem[{{Abramenko} \& {Longcope}(2005)}]{Abramenko2005}
{Abramenko}, V.~I., \& {Longcope}, D.~W. 2005, \apj, 619, 1160

\bibitem[{{Barthol} {et~al.}(2010){Barthol}, {Gandorfer}, {Solanki},
  {Sch{\"u}ssler}, {Chares}, {Curdt}, {Deutsch}, {Feller}, {Germerott},
  {Grauf}, {Heerlein}, {Hirzberger}, {Kolleck}, {Meller}, {M{\"u}ller},
  {Riethm{\"u}ller}, {Tomasch}, {Kn{\"o}lker}, {Lites}, {Card}, {Elmore},
  {Fox}, {Lecinski}, {Nelson}, {Summers}, {Watt}, {Mart{\'{\i}}nez Pillet},
  {Bonet}, {Schmidt}, {Berkefeld}, {Title}, {Domingo}, {Gasent Blesa}, {del
  Toro Iniesta}, {L{\'o}pez Jim{\'e}nez}, {{\'A}lvarez-Herrero},
  {Sabau-Graziati}, {Widani}, {Haberler}, {H{\"a}rtel}, {Kampf}, {Levin},
  {P{\'e}rez Grande}, {Sanz-Andr{\'e}s}, \& {Schmidt}}]{barthol_etal_2010}
{Barthol}, P., {et~al.} 2010, ArXiv e-prints

\bibitem[{{Bogdan} {et~al.}(1988){Bogdan}, {Gilman}, {Lerche}, \&
  {Howard}}]{Bogdan1988}
{Bogdan}, T.~J., {Gilman}, P.~A., {Lerche}, I., \& {Howard}, R. 1988, \apj,
  327, 451

\bibitem[{{Brandt} {et~al.}(1991){Brandt}, {Ferguson}, {Shine}, {Tarbell}, \&
  {Scharmer}}]{brandt1991}
{Brandt}, P.~N., {Ferguson}, S., {Shine}, R.~A., {Tarbell}, T.~D., \&
  {Scharmer}, G.~B. 1991, \aap, 241, 219

\bibitem[{{Brandt} {et~al.}(1988){Brandt}, {Scharmer}, {Ferguson}, {Shine}, \&
  {Tarbell}}]{brandt1988}
{Brandt}, P.~N., {Scharmer}, G.~B., {Ferguson}, S., {Shine}, R.~A., \&
  {Tarbell}, T.~D. 1988, \nat, 335, 238

\bibitem[{{Cattaneo} {et~al.}(2001){Cattaneo}, {Lenz}, \&
  {Weiss}}]{cattaneo2001}
{Cattaneo}, F., {Lenz}, D., \& {Weiss}, N. 2001, \apjl, 563, L91

\bibitem[{{de Wijn} {et~al.}(2005){de Wijn}, {Rutten}, {Haverkamp}, \&
  {S{\"u}tterlin}}]{de_wijn_etal_2008}
{de Wijn}, A.~G., {Rutten}, R.~J., {Haverkamp}, E.~M.~W.~P., \&
  {S{\"u}tterlin}, P. 2005, \aap, 441, 1183

\bibitem[{{Dom{\'{\i}}nguez Cerde{\~n}a}(2003)}]{dominguez2003}
{Dom{\'{\i}}nguez Cerde{\~n}a}, I. 2003, \aap, 412, L65

\bibitem[{{Dom{\'{\i}}nguez Cerde{\~n}a} {et~al.}(2003){Dom{\'{\i}}nguez
  Cerde{\~n}a}, {S{\'a}nchez Almeida}, \& {Kneer}}]{dominguezetal2003}
{Dom{\'{\i}}nguez Cerde{\~n}a}, I., {S{\'a}nchez Almeida}, J., \& {Kneer}, F.
  2003, \aap, 407, 741

\bibitem[{{Ishikawa} \& {Tsuneta}(2010)}]{ishikawa2010}
{Ishikawa}, R., \& {Tsuneta}, S. 2010, \apjl, 718, L171

\bibitem[{{Khomenko} {et~al.}(2003){Khomenko}, {Collados}, {Solanki}, {Lagg},
  \& {Trujillo Bueno}}]{khomenko_etal_2003}
{Khomenko}, E.~V., {Collados}, M., {Solanki}, S.~K., {Lagg}, A., \& {Trujillo
  Bueno}, J. 2003, \aap, 408, 1115

\bibitem[{{Leitzinger} {et~al.}(2005){Leitzinger}, {Brandt}, {Hanslmeier},
  {P{\"o}tzi}, \& {Hirzberger}}]{leitzinger2005}
{Leitzinger}, M., {Brandt}, P.~N., {Hanslmeier}, A., {P{\"o}tzi}, W., \&
  {Hirzberger}, J. 2005, \aap, 444, 245

\bibitem[{{Mart{\'{\i}}nez Gonz{\'a}lez} {et~al.}(2010){Mart{\'{\i}}nez
  Gonz{\'a}lez}, {Manso Sainz}, {Asensio Ramos}, \& {Bellot
  Rubio}}]{martinez_gonzalez2010}
{Mart{\'{\i}}nez Gonz{\'a}lez}, M.~J., {Manso Sainz}, R., {Asensio Ramos}, A.,
  \& {Bellot Rubio}, L.~R. 2010, \apjl, 714, L94

\bibitem[{{Martinez Pillet} {et~al.}(2010){Martinez Pillet}, {del Toro
  Iniesta}, {Alvarez-Herrero}, {Domingo}, {Bonet}, {Gonzalez Fernandez}, {Lopez
  Jimenez}, {Pastor}, {Gasent Blesa}, {Mellado}, {Piqueras}, {Aparicio},
  {Balaguer}, {Ballesteros}, {Belenguer}, {Bellot Rubio}, {Berkefeld},
  {Collados}, {Deutsch}, {Feller}, {Girela}, {Grauf}, {Heredero}, {Herranz},
  {Jeronimo}, {Laguna}, {Meller}, {Menendez}, {Morales}, {Orozco Suarez},
  {Ramos}, {Reina}, {Ramos}, {Rodriguez}, {Sanchez}, {Uribe-Patarroyo},
  {Barthol}, {Gandorfer}, {Knoelker}, {Schmidt}, {Solanki}, \& {Vargas
  Dominguez}}]{martinezpillet_etal_2010}
{Martinez Pillet}, V., {et~al.} 2010, ArXiv e-prints

\bibitem[{{Matloch} {et~al.}(2009){Matloch}, {Cameron}, {Schmitt}, \&
  {Sch{\"u}ssler}}]{matloch2009}
{Matloch}, L., {Cameron}, R., {Schmitt}, D., \& {Sch{\"u}ssler}, M. 2009, \aap,
  504, 1041

\bibitem[{{Matloch} {et~al.}(2010){Matloch}, {Cameron}, {Shelyag}, {Schmitt},
  \& {Sch{\"u}ssler}}]{matloch2010}
{Matloch}, {\L}., {Cameron}, R., {Shelyag}, S., {Schmitt}, D., \&
  {Sch{\"u}ssler}, M. 2010, \aap, 519, A52

\bibitem[{{Muller} {et~al.}(1992){Muller}, {Auffret}, {Roudier}, {Vigneau},
  {Simon}, {Frank}, {Shine}, \& {Title}}]{muller1992}
{Muller}, R., {Auffret}, H., {Roudier}, T., {Vigneau}, J., {Simon}, G.~W.,
  {Frank}, Z., {Shine}, R.~A., \& {Title}, A.~M. 1992, \nat, 356, 322

\bibitem[{{Nordlund} {et~al.}(2009){Nordlund}, {Stein}, \&
  {Asplund}}]{nordlund2009}
{Nordlund}, {\AA}., {Stein}, R.~F., \& {Asplund}, M. 2009, Living Reviews in
  Solar Physics, 6, 2

\bibitem[{{November} {et~al.}(1981){November}, {Toomre}, {Gebbie}, \&
  {Simon}}]{november1981}
{November}, L.~J., {Toomre}, J., {Gebbie}, K.~B., \& {Simon}, G.~W. 1981,
  \apjl, 245, L123

\bibitem[{{Orozco Su{\'a}rez} {et~al.}(2007){Orozco Su{\'a}rez}, {Bellot
  Rubio}, {del Toro Iniesta}, {Tsuneta}, {Lites}, {Ichimoto}, {Katsukawa},
  {Nagata}, {Shimizu}, {Shine}, {Suematsu}, {Tarbell}, \&
  {Title}}]{orozco_suarez_etal_hinode_magnetic_elements_2007}
{Orozco Su{\'a}rez}, D., {et~al.} 2007, \apjl, 670, L61

\bibitem[{{Rast}(2003)}]{rast2003}
{Rast}, M.~P. 2003, \apj, 597, 1200

\bibitem[{{Rieutord} {et~al.}(2001){Rieutord}, {Roudier}, {Ludwig}, {Nordlund},
  \& {Stein}}]{rieutord2001}
{Rieutord}, M., {Roudier}, T., {Ludwig}, H., {Nordlund}, {\AA}., \& {Stein}, R.
  2001, \aap, 377, L14

\bibitem[{{Rieutord} {et~al.}(2000){Rieutord}, {Roudier}, {Malherbe}, \&
  {Rincon}}]{rieutord2000}
{Rieutord}, M., {Roudier}, T., {Malherbe}, J.~M., \& {Rincon}, F. 2000, \aap,
  357, 1063

\bibitem[{{Roudier} {et~al.}(2003){Roudier}, {Ligni{\`e}res}, {Rieutord},
  {Brandt}, \& {Malherbe}}]{roudier2003}
{Roudier}, T., {Ligni{\`e}res}, F., {Rieutord}, M., {Brandt}, P.~N., \&
  {Malherbe}, J.~M. 2003, \aap, 409, 299

\bibitem[{{Roudier} {et~al.}(1998){Roudier}, {Malherbe}, {Vigneau}, \&
  {Pfeiffer}}]{roudier1998}
{Roudier}, T., {Malherbe}, J.~M., {Vigneau}, J., \& {Pfeiffer}, B. 1998, \aap,
  330, 1136

\bibitem[{{Roudier} \& {Muller}(2004)}]{roudier2004}
{Roudier}, T., \& {Muller}, R. 2004, \aap, 419, 757

\bibitem[{{Roudier} {et~al.}(2009){Roudier}, {Rieutord}, {Brito}, {Rincon},
  {Malherbe}, {Meunier}, {Berger}, \& {Frank}}]{roudier2009}
{Roudier}, T., {Rieutord}, M., {Brito}, D., {Rincon}, F., {Malherbe}, J.~M.,
  {Meunier}, N., {Berger}, T., \& {Frank}, Z. 2009, \aap, 495, 945

\bibitem[{{Roudier} {et~al.}(1999){Roudier}, {Rieutord}, {Malherbe}, \&
  {Vigneau}}]{roudier1999}
{Roudier}, T., {Rieutord}, M., {Malherbe}, J.~M., \& {Vigneau}, J. 1999, \aap,
  349, 301

\bibitem[{{Ruiz Cobo} \& {del Toro Iniesta}(1992)}]{ruiz_cobo1992}
{Ruiz Cobo}, B., \& {del Toro Iniesta}, J.~C. 1992, \apj, 398, 375

\bibitem[{{S{\'a}nchez Almeida}(2003)}]{sanchez_almeida2003}
{S{\'a}nchez Almeida}, J. 2003, \aap, 411, 615

\bibitem[{{S{\'a}nchez Almeida} {et~al.}(2010){S{\'a}nchez Almeida}, {Bonet},
  {Viticchi{\'e}}, \& {Del Moro}}]{sanchezalmeida_etal_2010}
{S{\'a}nchez Almeida}, J., {Bonet}, J.~A., {Viticchi{\'e}}, B., \& {Del Moro},
  D. 2010, \apjl, 715, L26

\bibitem[{{Seehafer}(1978)}]{seehafer1978}
{Seehafer}, N. 1978, \solphys, 58, 215

\bibitem[{{Shine} {et~al.}(2000){Shine}, {Simon}, \& {Hurlburt}}]{shine2000}
{Shine}, R.~A., {Simon}, G.~W., \& {Hurlburt}, N.~E. 2000, \solphys, 193, 313

\bibitem[{{Simon} {et~al.}(1988){Simon}, {Title}, {Topka}, {Tarbell}, {Shine},
  {Ferguson}, {Zirin}, \& {SOUP Team}}]{simon1988}
{Simon}, G.~W., {Title}, A.~M., {Topka}, K.~P., {Tarbell}, T.~D., {Shine},
  R.~A., {Ferguson}, S.~H., {Zirin}, H., \& {SOUP Team}. 1988, \apj, 327, 964

\bibitem[{{Solanki} {et~al.}(2010){Solanki}, {Barthol}, {Danilovic}, {Feller},
  {Gandorfer}, {Hirzberger}, {Riethm{\"u}ller}, {Sch{\"u}ssler}, {Bonet},
  {Mart{\'{\i}}nez Pillet}, {del Toro Iniesta}, {Domingo}, {Palacios},
  {Kn{\"o}lker}, {Bello Gonz{\'a}lez}, {Berkefeld}, {Franz}, {Schmidt}, \&
  {Title}}]{solanki_etal_2010}
{Solanki}, S.~K., {et~al.} 2010, \apjl, 723, L127

\bibitem[{{Welsh} {et~al.}(2004){Welsh}, {Fisher}, {Abbett}, \&
  {Regnier}}]{welsh2004}
{Welsh}, B.~T., {Fisher}, G.~H., {Abbett}, W.~P., \& {Regnier}, S. 2004, \apj,
  610, 1148

\bibitem[{{Wiegelmann}(2004)}]{wiegelmann2004}
{Wiegelmann}, T. 2004, \solphys, 219, 87

\bibitem[{{Wiegelmann} {et~al.}(2010{\natexlab{a}}){Wiegelmann}, {Yelles
  Chaouche}, {Solanki}, \& {Lagg}}]{wiegelmann2010a}
{Wiegelmann}, T., {Yelles Chaouche}, L., {Solanki}, S.~K., \& {Lagg}, A.
  2010{\natexlab{a}}, \aap, 511, A4

\bibitem[{{Wiegelmann} {et~al.}(2010{\natexlab{b}}){Wiegelmann}, {Solanki},
  {Borrero}, {Mart{\'{\i}}nez Pillet}, {del Toro Iniesta}, {Domingo}, {Bonet},
  {Barthol}, {Gandorfer}, {Kn{\"o}lker}, {Schmidt}, \&
  {Title}}]{wiegelmann2010b}
{Wiegelmann}, T., {et~al.} 2010{\natexlab{b}}, \apjl, 723, L185

\end{thebibliography}
\end{document}